\documentclass[letter,12pt]{article}
\usepackage{graphicx}
\begin{document}

\title{Continuous opinion model in small world directed networks}
\author{Y\'{e}rali Gandica $^{1}$\\
\emph{Instituto Venezolano de Investigaciones Cient\'{\i}ficas.}\\
\emph{Centro de F\'{\i}sica, Altos de Pipe, Carretera Panamericana, Km 11,}\\
\emph{Caracas 1020A, Venezuela.} \and Marcelo del Castillo-Mussot,
Gerardo J. V\'{a}zquez. \\
\emph{Departamento de Estado S\'{o}lido, Instituto de F\'{\i}sica,}\\
\emph{Universidad Nacional Aut\'{o}noma de} \emph{M\'{e}xico, Apdo. Postal
20-364,}\\
\emph{01000 M\'{e}xico, D.F., M\'{e}xico} \and Sergio Rojas. \\
\emph{Departamento de F\'{\i}sica, Universidad Sim\'{o}n Bol\'{\i}var,}\\
\emph{Valle de Sartenejas, Baruta, }\\
\emph{Apartado Postal 89000, Caracas 1080-A, Venezuela.}}
\maketitle

\begin{abstract}
In the compromise model of continuous opinions proposed by Deffuant et al,
the states of two agents in a network can start to converge if they are
neighbors and if their opinions are sufficiently close to each other, below
a given threshold of tolerance $\epsilon$. In directed networks, if agent i is a
neighbor of agent j, j need not be a neighbor of i. In Watts-Strogatz
networks we performed simulations to find the averaged number of final
opinions $<F>$ and their distribution as a function
of $\epsilon$ and of the network structural disorder. In directed networks 
$<F>$ exhibits a rich structure, being larger than
in undirected networks for higher values of $\epsilon$, and smaller for lower
values of $\epsilon$.
\textit{Keywords}: Opinions Dynamics, Convergence Pattern,
Consensus/dissent, bounded confidence, directed networks.

\end{abstract}

\footnotetext{$^{1}$Correspondence author. E-mail: ygandica@gmail.com}
\newpage 
\baselineskip16pt plus 2pt minus 1pt
\section{INTRODUCTION}

In many social, biological, and economic systems there are complex networks
that include directed links acting only in one direction; outwards or
inwards. That is, there are processes in which agents do not act
in a symmetrical way. Therefore, studying and comparing different dynamics
in similar undirected (UN) and directed networks (DN) is an important task.
In social systems such as families, clans, schools, etc. there exist
directed or asymmetrical processes like copying or imitating, giving orders,
teaching, etc. Similar mechanisms occur in automatized or control systems
regarding flux of information or signals.

We can visualize asymmetric cases with arrows pointing in only one direction
to indicate the lack of reciprocity or bilaterality. Only one direction 
instead of two is equivalent to cutting links. Therefore, when some agents 
are chosen in a network, it is possible that they remain totally passive, 
i. e. do not change their state.

Our networks models are based on the Watts-Strogatz structure \cite{Watts/Stogatz} 
for small-world phenomena, where in DN there is only one 
link instead of bidirectional double links connecting the same two vertices 
or nodes in the corresponding UN.

We will now briefly mention some representative works on directed networks.
Sanchez \textit{et al} \cite{Sanchez} investigated the effect of directed
links on the behavior of a simple spin-like model evolving on a small-world
network, leading to a phase diagram including first- and second-order phase
transitions out of equilibrium. The majority-vote model has been studied
with noise on directed random graphs \cite{Lima}. Inspired in food web
theory in ecosystems, Morelli \cite{Morelli} stu\-died the fraction of basal,
top nodes and node level distributions in directed networks based on the
Watts-Strogatz model, gave analytical expressions for the fraction of basal
and top nodes for the model, and studied the node level distributions with
numerical simulations. When a naive spreading process starts in a directed
network, the interplay of shortcuts and unfavorable bonds on the small world
properties is studied in simple models of small-world networks with directed
links \cite{Ramez}, leading to general results small-world networks with
directed links.

On the other hand, convergence (or divergence) of ideas or opinions among
participants of a debate is a very important social process. In practice, in
opinion models it is plausible to assume that discussions take place when
the opinions of the people involved are sufficiently close to each other, a
process called bounded confidence. Then agents can negotiate their
difference to try to reach consensus, or al least, to share similar, if not
equal, opinions. In Deffuant \textit{et al }model \cite{Deffuant} (hereafter
referred to, for simplicity, as Deffuant model), a simple opinion dynamics
was proposed for the full graph, which we employ here in both UN and DN
within the Watts-Strogatz model. Some models of opinion dynamics, including
Deffuant model, have been reviewed in Lorenz \cite{Lorenz} and Castellano 
\textit{et al} \cite{Castellano}. We now mention some papers based on this
model. The effect of varying the number of peers met at one time, for
different population sizes, and the effects of changing the self-support in
the Deffuant model was studied by Urbig \textit{et al} \cite{Urbig}. Huet 
\textit{et al} \cite{Huet}\ added a rejection mechanism into a 2D bounded
confidence model based in the Deffuant model, yielding metastable clusters,
which maintain themselves through opposite influences of competitor
clusters. The Deffuant model has been studied also in scale free network
topology\cite{Guo,Weisbuch,Jacobmeier,Stauffer,Fortunato}. Guo and Cai 
\cite{Guo} took into account the heterogeneous distribution of connectivity
degree in an Deffuant model in which the convergence parameter of an agent
is a function of the degree of its interaction partner, and in particular,
they assumed that convergence is faster in agents interacting with more
connected or ``famous'' agents (``celebrity'' effect).  Within
the Deffuant model Groeber \textit{et al} \cite{Groeber} combined an agent's
behavior and the mean behaviour of her in-group (agent's past interaction
partners in her neighborhood) to foster consensus and to yield new local
clusters or cultures. Within the same spirit of incorporating the states of
each agents neighborhood, the transition from invasion to coexistence in nonlinear
voter models has been investigated \cite{Schweitzer} to yield three regimes:
complete invasion; random coexistence; and correlated coexistence. In Ref. \cite{Shao}, 
a node will convert to its opposite opinion, if it is in
the local minority opinion, but in contrast with the majority-voter
model, the opinion of each node itself is included with its
neighbors. Then, because of the clustering (community support) of agents
holding the same opinion, these clusters cannot be invaded by the other opinion, 
analogously to incompressible fluids, a fact that allows to map this opinion clustering 
behavior to a known physics percolation problem \cite{Schwarzer}. In a
discrete Deffuant model on a directed Barabasi-Albert, it is shown in \cite{Jacobmeier} 
that it is difficult to reach absolute consensus. In a similar
network \cite{Stauffer}, a multi-layer model representing various age levels
was employed and advertising effects were included. In Deffuant model,
extremists were defined \cite{deff-jasss} as individuals with a very low
uncertainty in their opinion states, and located at the extremes of the
initial opinion distribution. In \cite{Frederic}, the role of network topology on extremism
propagation in small-world networks was investigated and a drift to a single
extreme appeared only beyond a critical level of connectivity, which
decreases when the randomness increases. Assmann \cite{Assmann} investigated
agents with different random and systematic opinion qualities in the
Deffuant model.

In this paper, we investigate the effect of directed links on the behavior
of a simple continuous opinion model, the Deffuant model, evolving on
disordered networks. Our simulations could help understand opinion
formation in networks with less number of links due to lack of reciprocity
and disordered heterogeneity. After reviewing the two main elements of our
model; the dynamics of Deffuant model and secondly, the directed networks,
here we combine them to simulate the Deffuant model in $directed$
Watts-Strogatz networks for small-world phenomena. In Sect. II we present
the general opinion dynamics in Deffuant model and the construction of the
directed network (topology). Sect. III is devoted to results and discussion
of the differences of the Deffuant model in UN and DN regarding clustering,
convergence time, distribution of final opinions and conservation of
opinion. Sect. IV we present our summary and conclusions.

\section{THE MODEL}

In the Deffuant model\textit{\ }\cite{Deffuant}, a simple opinion dynamics
was proposed for the fully connected graph, where all nodes are connected
among them (or any agent or node has all nodes as neighbors). Here we apply
the notion of neighbors connected by (bi-directional or uni-directional)
links. In the case of directed networks $j$ is neighbor of $i$ only if there
exist a link from $i$ to $j$. The dynamic is summarized as follows: On any
issue, each opinion is represented by a continuous number $x$
chosen between $0.0$ and $1.0$, without loss of generality. One selects an
agent $i$ and then with the same probability for all its neighbors \ (if
there are any) one selects one of them called $j$. 

Then between the selected neighbors, if the difference of the opinions $%
x_{i}(t)$ and $x_{j}(t)$ exceeds the threshold, openness or tolerance, $%
\epsilon $, nothing happens (here $t$ is a \textit{dynamical} time that
labels and orders the time steps of the iterative process), but if, $%
|x_{i}(t)-x_{j}(t)|<\epsilon $, then:

\begin{eqnarray}
x_{i}(t+1) &=&x_{i}(t)+\mu \lbrack x_{j}(t)-x_{i}(t)],  \label{uno} \\
x_{j}(t+1) &=&x_{j}(t)+\mu \lbrack x_{i}(t)-x_{j}(t)].  \nonumber
\end{eqnarray}

which means that their opinions get closer, as measured by the
parameter $\mu$ $ ( \mu \in [0,\frac{1}{2}]),$
This process can be labeled as a ``negotiation''.

This procedure is repeated until all opinion values do not change, and then
it is said that convergence is reached. It is usual to use the same
approaching parameter $\mu $ for all agents. If $\mu $ = $\frac{1}{2}$ then
both opinions \ $x_{i}(t)$ and $x_{j}(t)$ become the same in one step at
their midpoint and consensus of the two agents involved is then very quick
and complete. Obviously the underlying topology of the system defines the
neighborhood of every agent.  When convergence is reached, after employing
iteratively the Eq. \ref{uno} procedure, we define a cluster as the set of
nodes sharing the same final opinions.

A common procedure to create a disordered network, known as the Watts $\&$
Strogatz model (WSM) \cite{Watts/Stogatz}, is to start from a regular
structure with \ $N$ \ nodes (in our case nodes on a circle connected with
nearest neighbors) of connectivity $k$, and then remove each link with
probability $p$, reconnecting it at random. The random rewiring process
introduces $pNk$ long links which connect nodes that otherwise were
connecting only nearest neighbors. The resulting network from the above
procedure is an undirected (or more pro\-per\-ly called in this case
bi-directional) network, because each link connects or acts in both
direcctions. For the DN case each bi-directional link is reemplaced by a
line with only one arrow, which direction is randomly selected.  Therefore, 
if directedness (or presence of arrows) is not taken into account, the directed 
network (DN) has exactly the same structure as the undirected one (UN). 

\section{Results and discussion}

\begin{figure}[ht]
\hbox{
\includegraphics[width=7.5 cm]{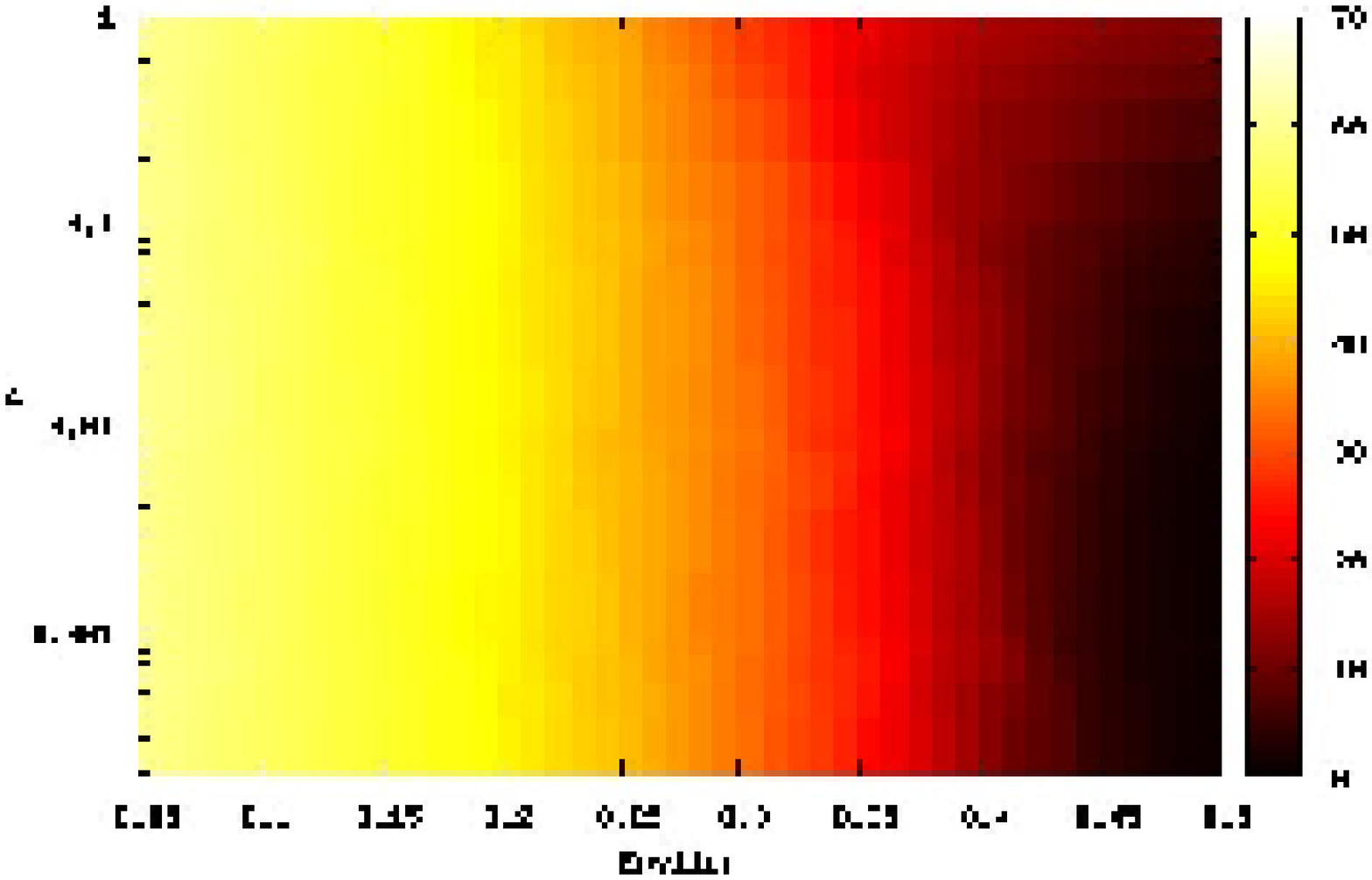} 
\includegraphics[width=7.5cm]{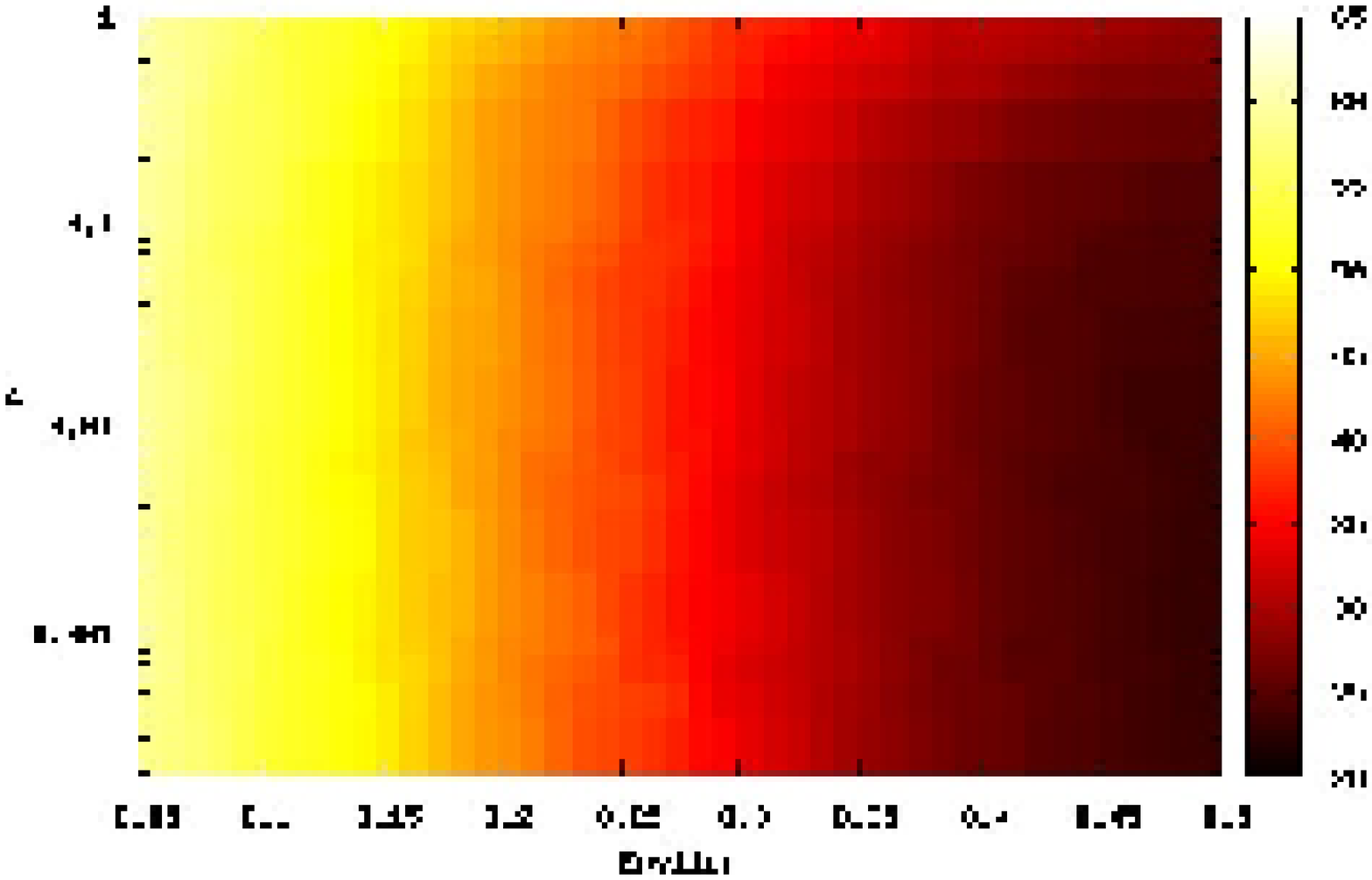}
     }
\caption{{Final opinions  averaged over $100$ realizations. $\left\langle F_{UN}\right\rangle $ (UN) on
the left panel and $\left\langle F_{DN}\right\rangle$ (DN) on the right panel. For $N=100$, $k=2$.}}
\label{fig1}
\end{figure}

\begin{figure}[ht]
\hbox{
\includegraphics[width=7.5 cm]{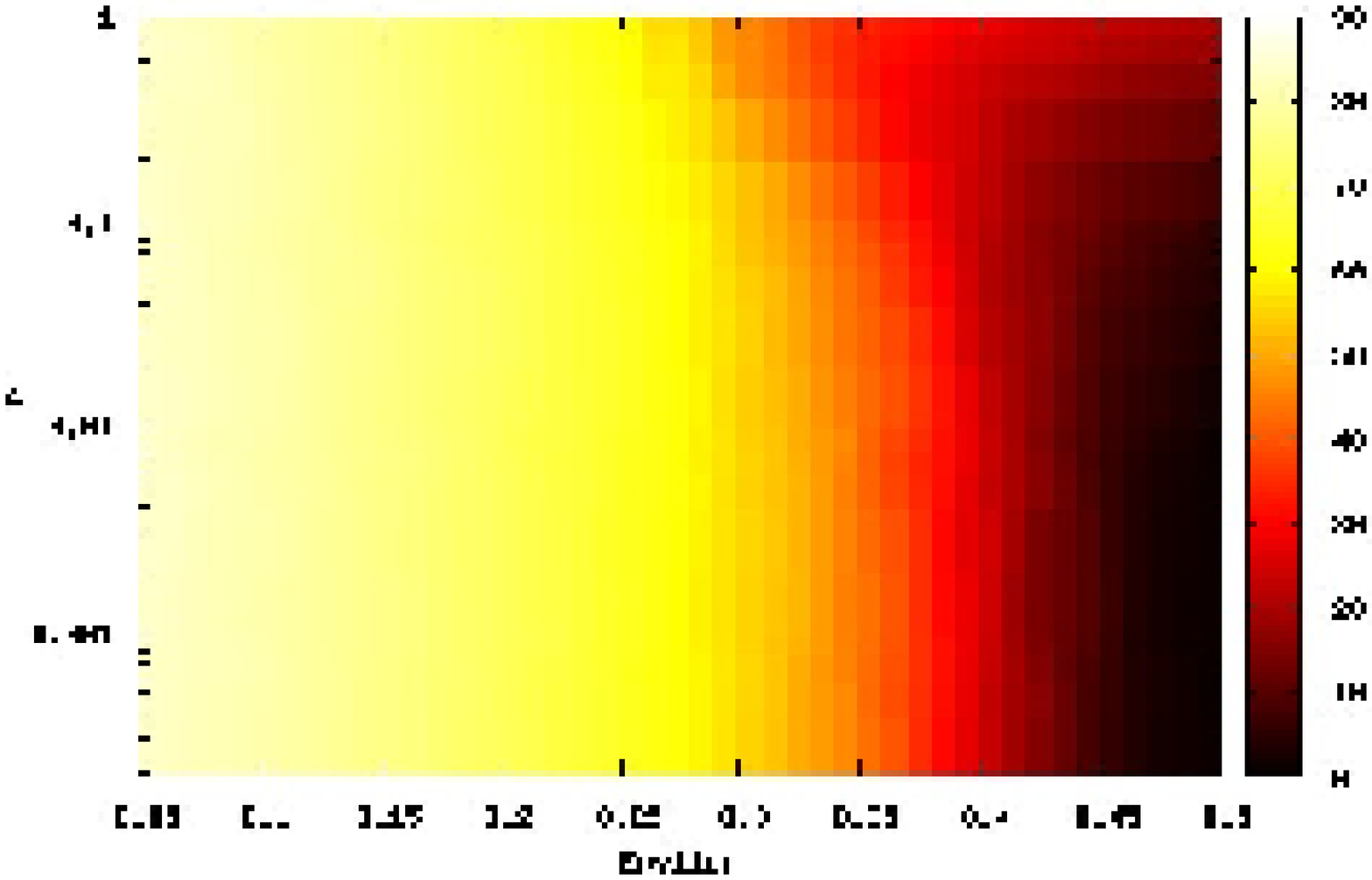} 
\includegraphics[width=7.5cm]{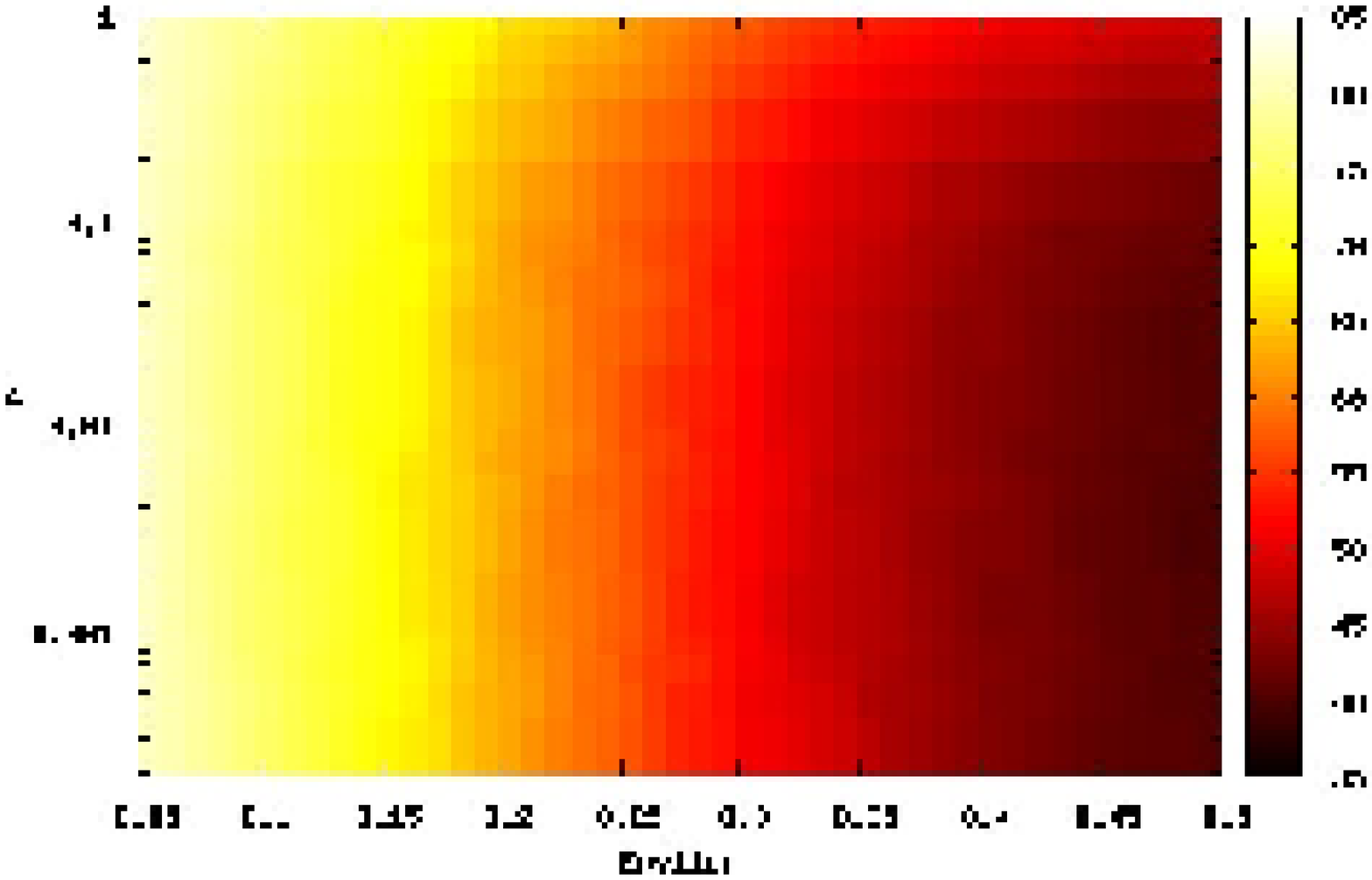}
     }
\caption{{Final opinions  averaged over $100$ realizations. $\left\langle F_{UN}\right\rangle $ on
the left panel and $\left\langle F_{DN}\right\rangle$ on the right panel. For $N=200$, $k=2$.}}
\label{fig2}
\end{figure}

\begin{figure}[ht]
\hbox{
\includegraphics[width=7.5 cm]{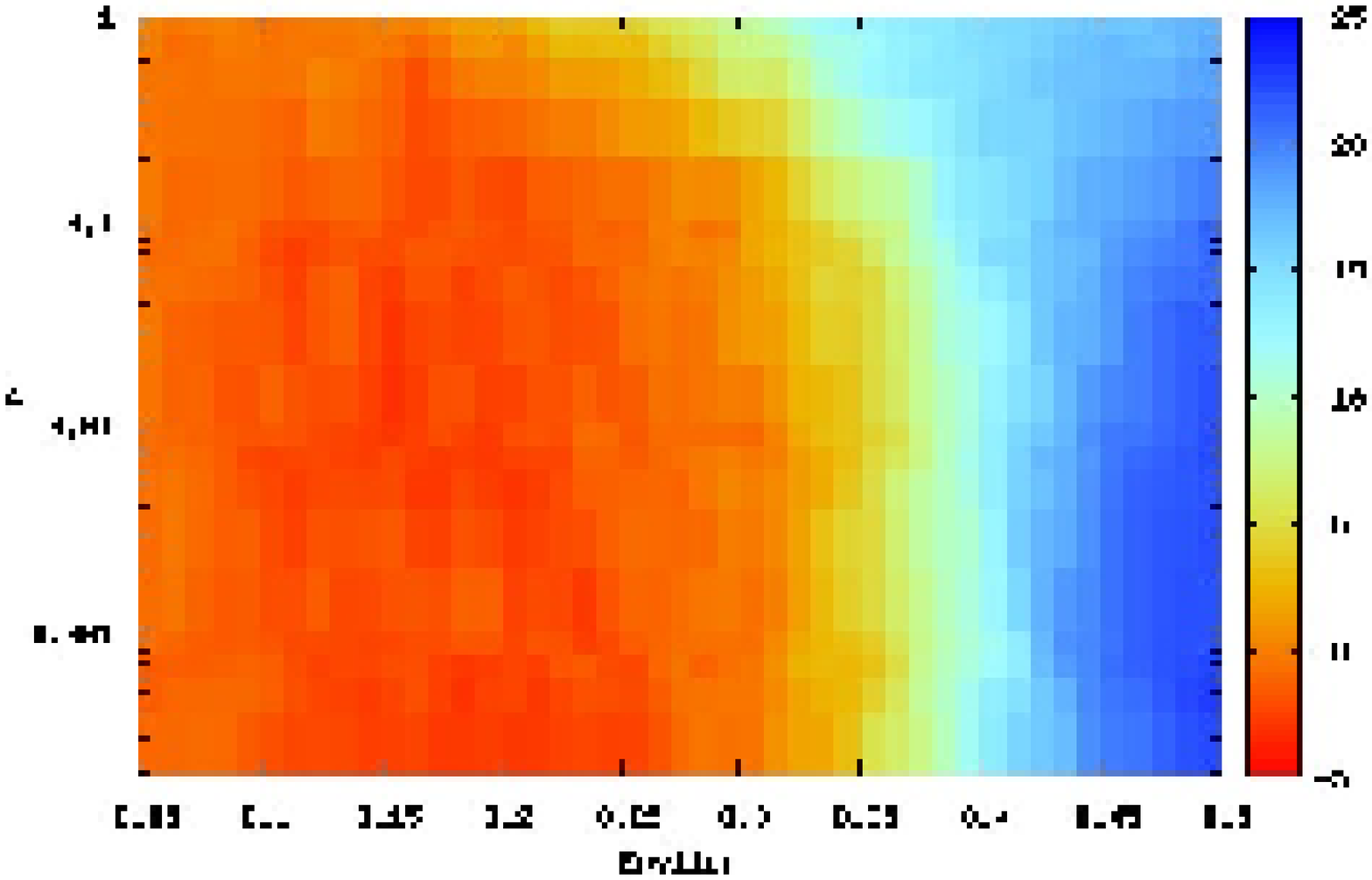} 
\includegraphics[width=7.5cm]{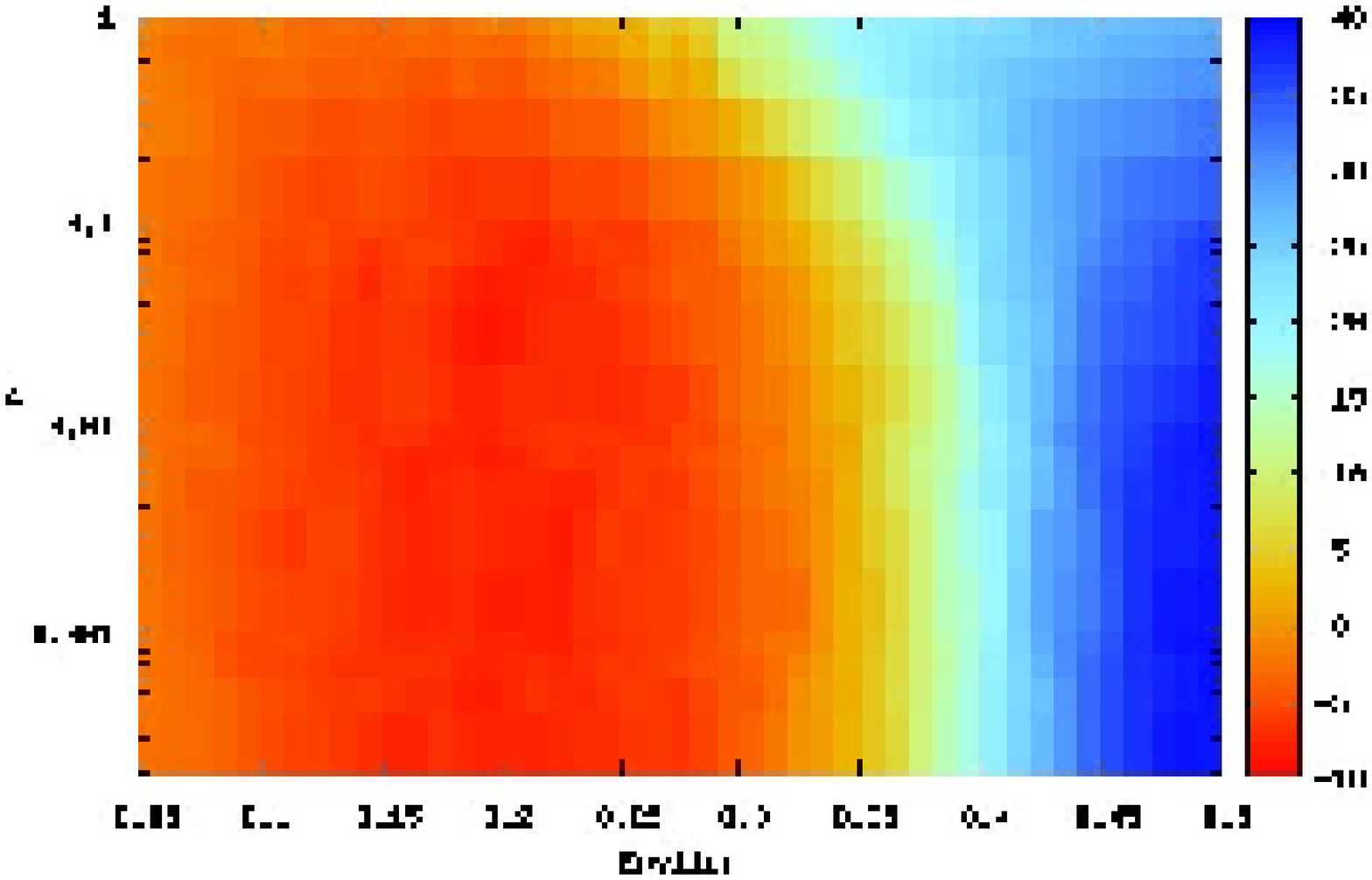}
}
\caption{{$\left\langle F_{DN}\right\rangle$ - $\left\langle F_{UN}\right\rangle$ . For $N=100$ on the left panel and $N=200$ on the right panel. Averaged over 100 realizations and $k=2$.}}
\label{fig3}
\end{figure}

In Figs.\ref{fig1} and \ref{fig2} we show the number of final states $%
\left\langle F\right\rangle $ for networks with $N=100$ and $N=200,$
respectively, averaged over $100$ realizations. We notice that in both UN
and DN, $\left\langle F\right\rangle $ exhibits a much stronger dependence
on the the tolerance $\epsilon $ than on the disorder parameter $p$. As in
any bounded confidence model, increasing $\epsilon $ leads to a monotonous
decrease in $\left\langle F\right\rangle $, an effect that is expected to be
more pronounced in the case of the UN, due to a larger (double) number of
links  that lead to a more refined convergence process. In contrast, there
is a very weak dependence of $\left\langle F\right\rangle $ on $p$ in both
UN and DN.
To compare the results of Figs.\ref{fig1} and \ref{fig2}, in Fig.\ref{fig3} we plot the
difference of $\ $final opinions $\left\langle F_{DN}\right\rangle
-\left\langle F_{UN}\right\rangle $ for $N=100$ and $N=200$, which present
the same qualitative behavior. As mentioned above, this difference as
function de $\epsilon $ is expected to be \ positive and large due to larger
number of links in the UN. This is the case for large values of $\epsilon ,$
but surprisingly, the difference exhibits small negative values for lower
values of $\epsilon $.

\begin{figure}[ht]
\hbox{
\includegraphics[width=7.5 cm]{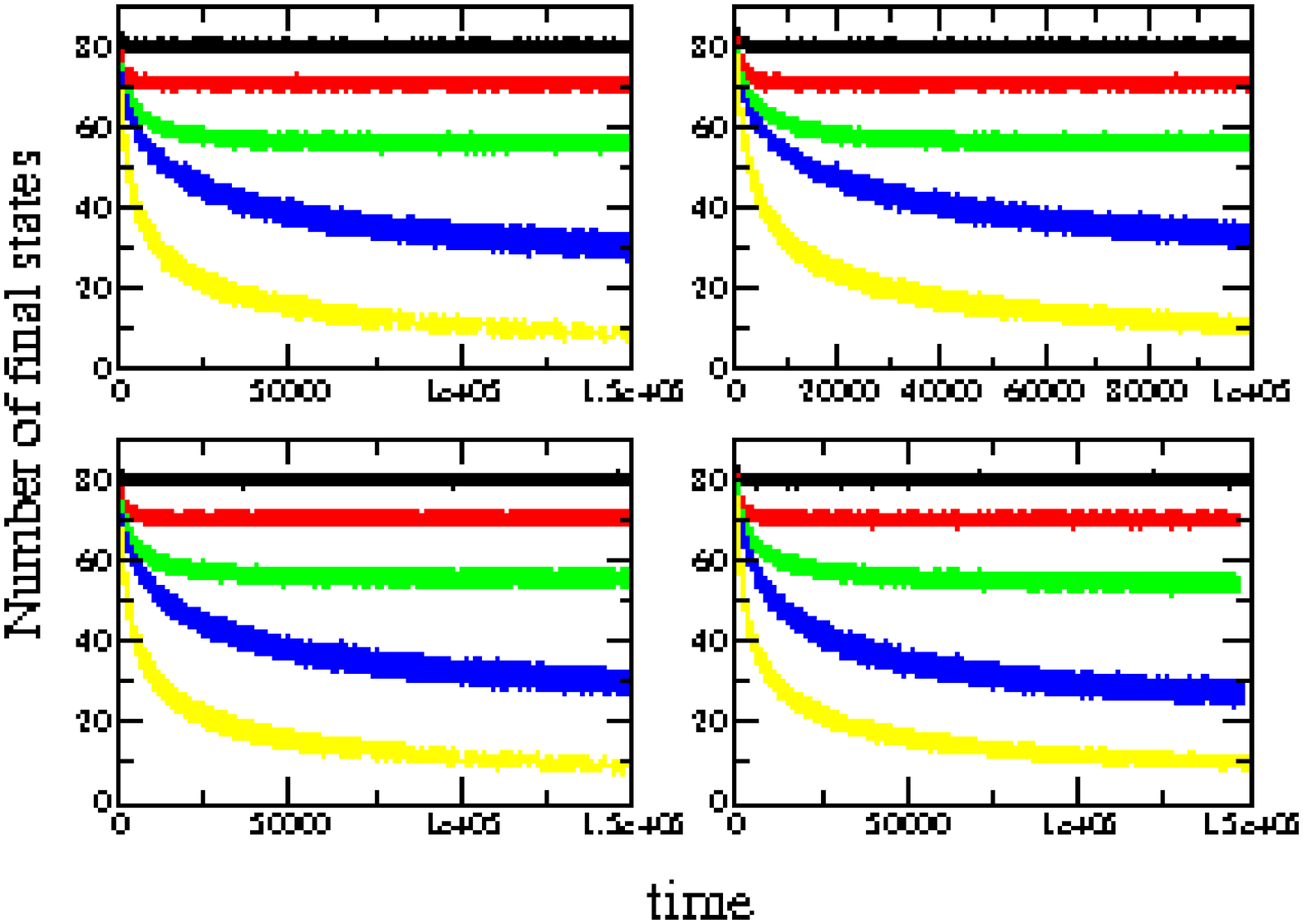}
\includegraphics[width=7.5 cm]{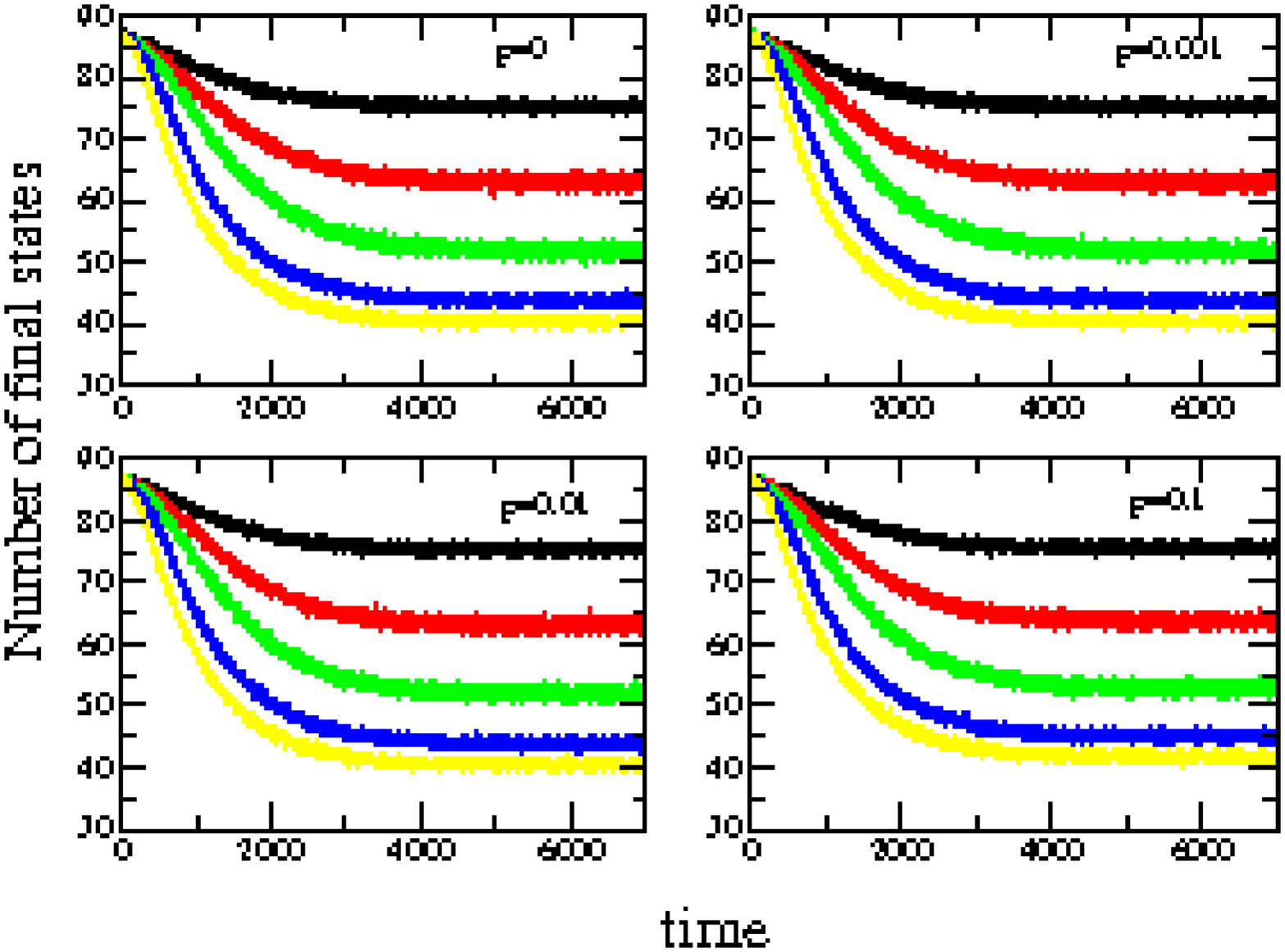}
}
\caption{\label{fig4}{Time evolution of the number of final opinions, for N=200, k=20. Averaged over 100 re\-a\-li\-za\-tions. UN on the left panel and DN on the right panel. For disorder values of: top left $p=0$, top right $p=0.001$, lower left $p=0.01$ and lower right $p=0.1$. In each graph black corresponds to $\epsilon=0.1$, red $\epsilon=0.2$, green $\epsilon=0.3$,
blue $\epsilon=0.4$ and yelow $\epsilon=0.5$.}}
\end{figure}

To understand this behaviour we get more detailed information by calculating
the time evolution of the system and distributions of the final opinions as
a function of $\epsilon $ \ and $p$. We plot in Figs. \ref{fig4} the time
evolution of the number of final opinions $\left\langle F\right\rangle $
averaged over 100 realizations for a set of different parameters. Here each
time unit is defined in terms of the system size as $N$ dynamical steps or
sequential loops of Eq. \ref{uno}. Notice that the convergence time is much
larger in the UN than in the DN, and this difference grows with $\epsilon $.
The presence of less number of links in the DN cause a much faster
convergence, whereas the presence of disorder affects little the convergence
time in both UN and DN.

\begin{figure}[ht]
\hbox{
\includegraphics[width=7.5 cm]{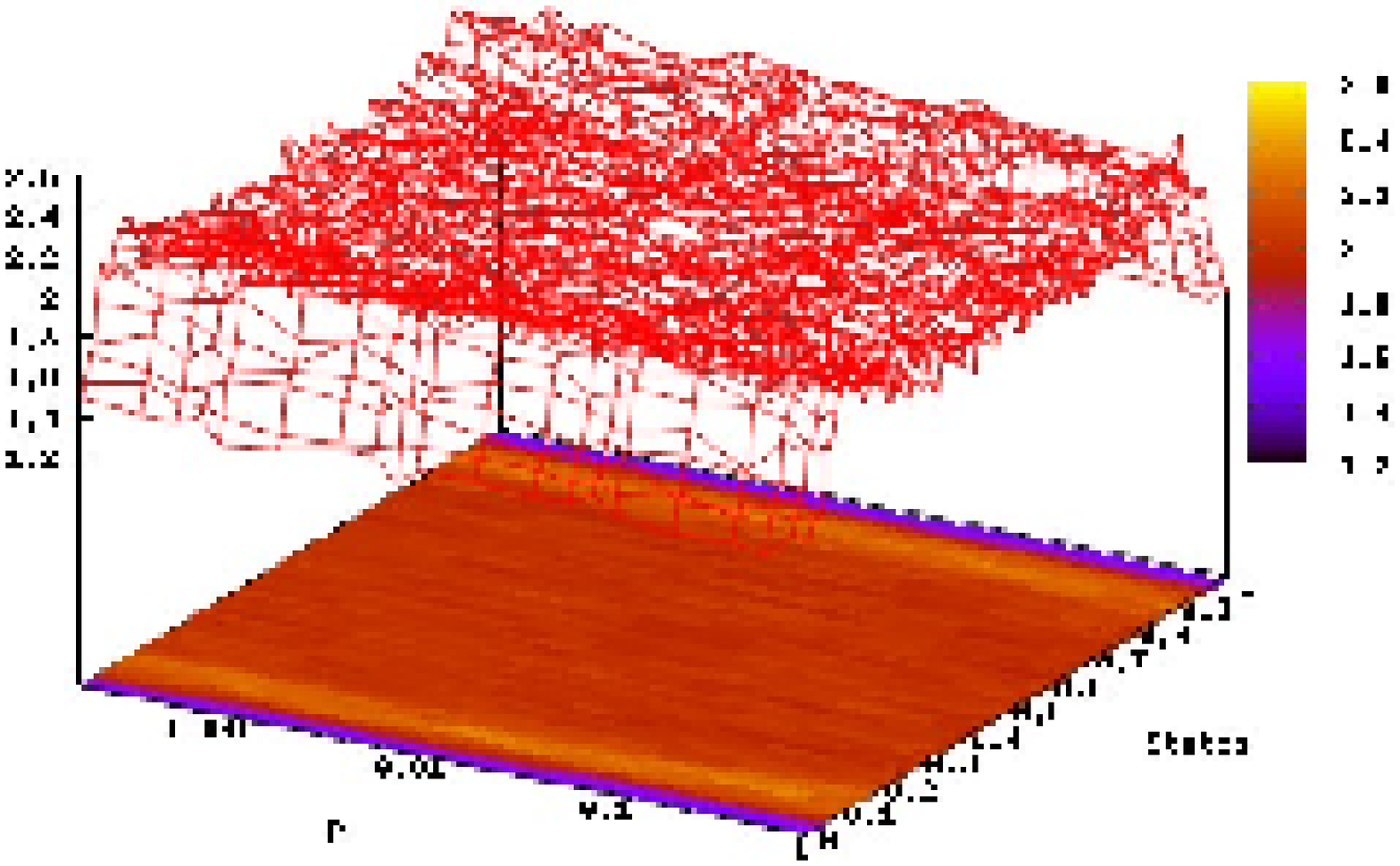} %
\includegraphics[width=7.5 cm]{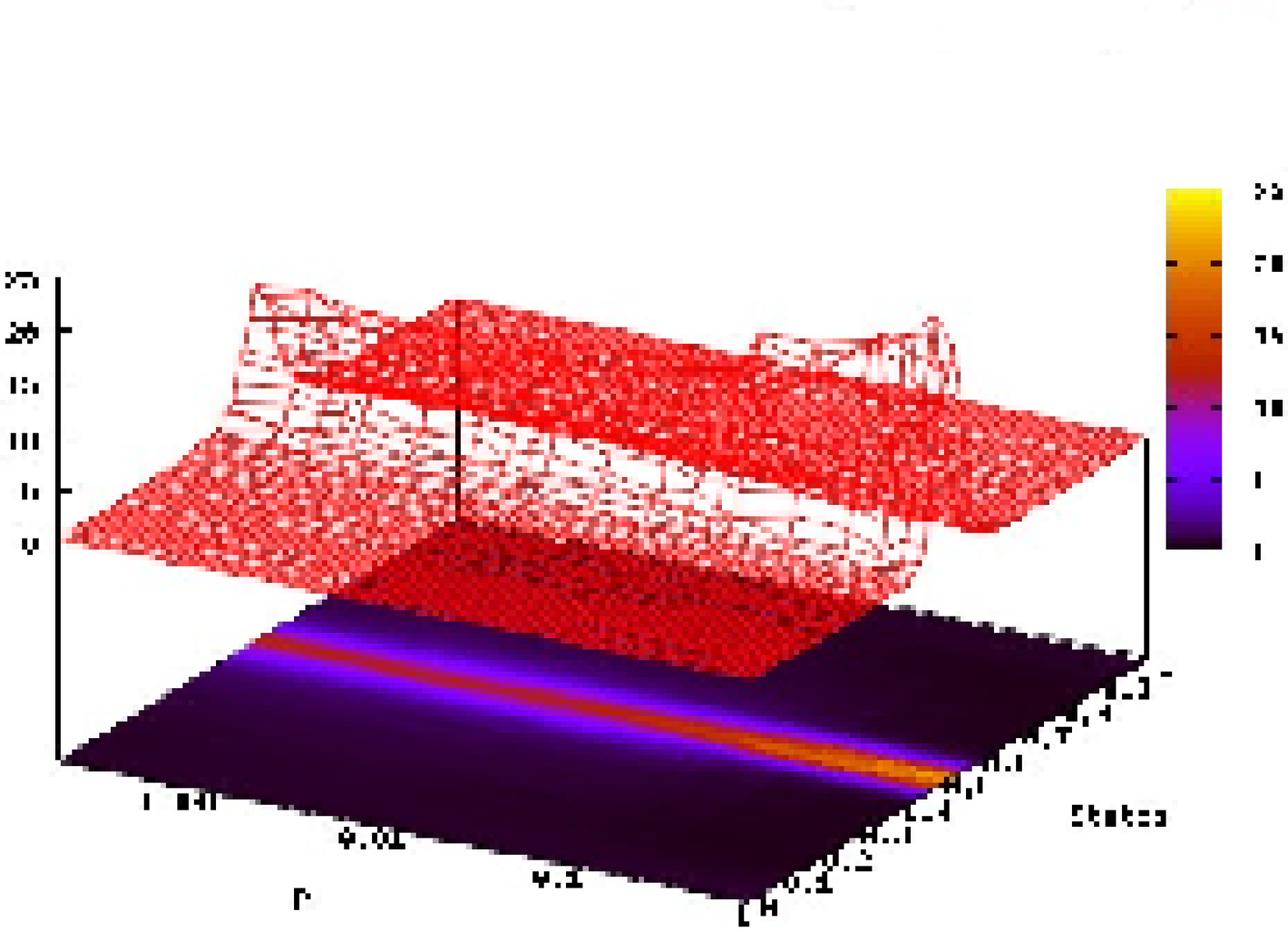}
}
\caption{{Distribution of final opinions for UN networks, N=200,
averaged over 100 re\-a\-li\-za\-tions. For $\protect\epsilon =0.15$ on left panel and $%
\protect\epsilon =0.4$ on right panel.}}
\label{fig5}
\end{figure}

\begin{figure}[ht]
\hbox{
\includegraphics[width=7.5 cm]{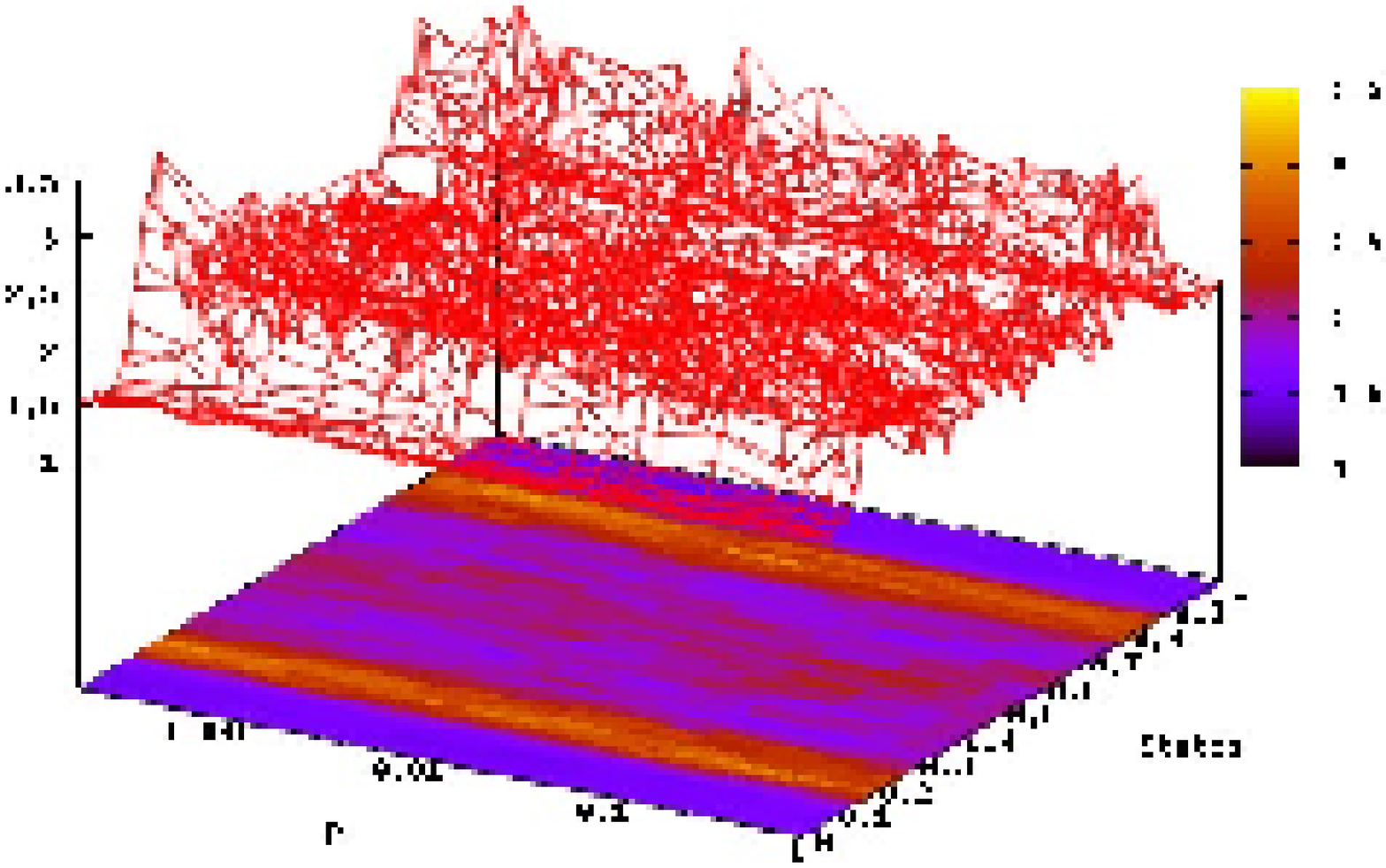} %
\includegraphics[width=7.5 cm]{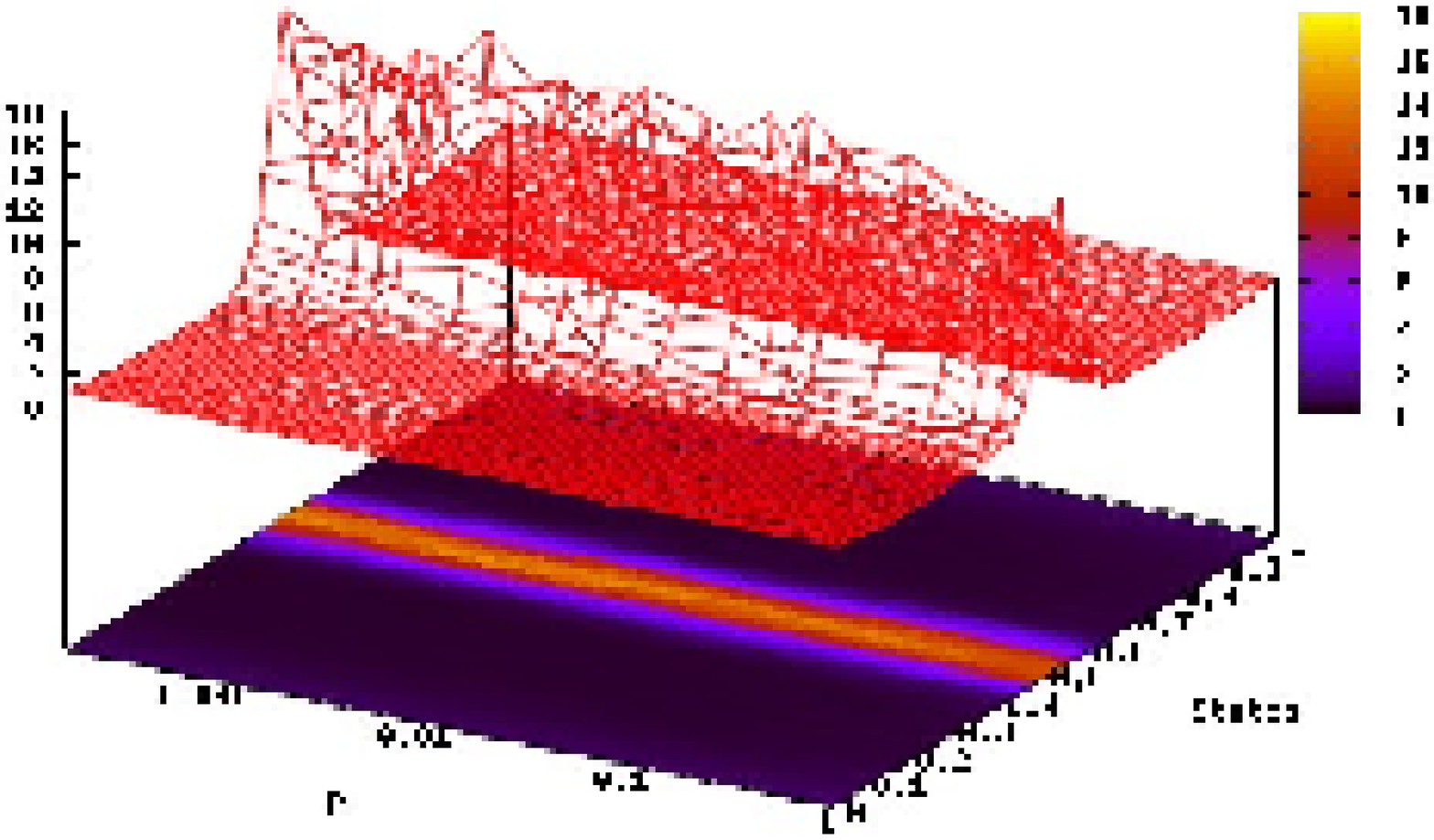}
}
\caption{{Distribution of final opinions for DN networks, N=200,
averaged over 100 re\-a\-li\-za\-tions. For $\protect\epsilon =0.15$ on left panel and $%
\protect\epsilon =0.4$ on right panel.}}
\label{fig6}
\end{figure}

In Fig. \ref{fig5} (UN) and Fig. \ref{fig6} (DN) we show examples of final
opinion distributions in the interval [0,1] for two different values of
tolerance; $\epsilon =0.15$ and $0.4$.


Results in UN exhibit a more smoother behavior as compared with those of the
DN (with larger number of links), in the sense that the DN curves show a
more regular and symmetrical behaviour in [0,1] due to slower and better
convergence. As expected, for high values of tolerance, the system converges
to only one cluster with the average value of all states, but surprisingly
for small values of the tolerance, clusters with opinion values closer to
the opinion edges 0 and 1 interact less than those near the middle, that
have neighbors with similar opinions at both their right and left flanks.
This effect is more pronounced in the DN, for which even some clusters are
lost in the middle of [0,1] for values of $\epsilon $ less than
0.3, approximately. If we roughly divide the interval [0,1] in thirds, it is
the middle section that tends to be emptied because there always exist
opinions at both right and left ranges, while opinions at the edge ``thirds''
are more restricted to interact with neighbors in  one side, rather than
in a less symmetric opinion space. Since in the DN a fewer number
of links yield much shorter convergence times, then nodes around the middle
in opinion space have less oportunities to come back to middle opinions
after interacting with agents with more extreme opinions. The fact that
clusters around the middle loose members or strength was already
observed in UN on the curves for the population of equally-spaced adjacent
opinion clusters \ or \textquotedblleft parties\textquotedblright\ by
Ben-Naim, \textit{et al } \cite{Ben-Naim}, who employed a density-based
dynamics for a density function to determine the agents density in the
opinion space. Here, besides this depopulating effect, we also found that in
the DN some final opinions do not survive at all, which explains the
negative values in $\left\langle F_{DN}\right\rangle -\left\langle
F_{UN}\right\rangle $ in Fig. \ref{fig4}

To summarize the dependence of our results on $\epsilon $, we emphasize that
for higher values of $\epsilon $ there are more nodes or agents in the DN
that are isolated or jammed that in the UN case. Then  the number of final
opinions $\left\langle F\right\rangle $ in DN is much larger, as compared
with the UN, because of the relative absence of links. That is in DN there
are more isolated agents or pathway-endings. The larger number of final
opinion in the DN resembles results in UN models that include many
extremists or \textquotedblleft stubborn\textquotedblright\ \cite{Frederic}
agents with small capacity to change opinion. Of course, the underlying
mechanisms are different; in the former case (DN), there are less links than
in the usual UN, and in the latter UN case agents behavior is not
homogeneous\ by construction. In the region of lower $\epsilon $, the fact
that $\left\langle F_{DN}\right\rangle -\left\langle F_{UN}\right\rangle $
is negative was explained in terms of the irregularity and
lost of clusters in the DN, relative to the UN.

The dependence of both $\left\langle F_{DN}\right\rangle $ and $\left\langle
F_{UN}\right\rangle $ on \ $p$ is more much smaller than on $\epsilon $.
Figs.\ref{fig1} and \ref{fig2} show how the increase of disorder (increasing
\ $p$) favors long-range dissemination and convergence in the case of $%
\left\langle F_{UN}\right\rangle $ and when $\epsilon $ also increases. In
general, it is known that there is a competition of local and global
dissemination in complex networks, since increasing disorder favours
long-range dissemination. The presence of a small fraction of
\textquotedblleft short cuts\textquotedblright\ connecting otherwise distant
points, drastically reduces the average shortest distance between any pair
of nodes in network, keeping the clustering high \cite{Ramez,Guo}. In our
dynamics, i.e. the Deffuant mode, the effect of the short cuts seems to be
less important than, for instance, in the Ising model in small-world
topologies, for which, when directed links are considered, not only the
critical temperature changes but the nature of the transition also switches%
\cite{Sanchez}. In the Axelrod model of cultural dissemination, the same
qualitative result as for the equilibrium Ising model was found, in the
sense that the small-world connectivity favors homogeneous or ordered states 
\cite{Klemm}

On the other hand, when $p$ increases and $\epsilon $ is large, $%
\left\langle F_{DN}\right\rangle $ may slightly decrease, in contrast to $%
\left\langle F_{UN}\right\rangle $, showing that short cuts are not so
efficient in the DN in favoring convergence in the Deffuant model. However,
the behavior of $\left\langle F_{DN}\right\rangle -\left\langle
F_{UN}\right\rangle $ is more similar to that of $\left\langle
F_{UN}\right\rangle $.

Finally, in our simulations we found that the value of the final opinions in
each cluster for both UN and DN was the $exact$ average of the initial
opinions for that cluster, regardless of the random process to reach the
final consensus within each final cluster $C$. In a similar direction,
Jacobmeier \cite{Jacob} employed the Deffuant model on a directed
Barabasi-Albert network with discrete opinions, and pointed out that the
resulting opinion distribution converged towards the a\-ve\-ra\-ge value of
the initial opinion distribution, which was identified as a guide of opinion
forming. Ben-Naim \textit{et al } \cite{Ben-Naim} also showed how opinion is
conserved in the particular case $\mu $ =$\frac{1}{2}$ with a density
formalism. Here, for any value $\mu ,$ we outline a proof of how the
conservation of opinion leads to clusters with the final average of the
initial opinion values for both UN and DN. Of course, this conservation
property is a consequence of the symmetrical approach process; that is, this
property is arises from the same values of the ``approaching'' parameter $\mu $
for both agents $i$ and $j$ in Eq. \ref{uno}.

To prove that all connected nodes will converge to the average of the
initial opinions, we start with the simple case $\epsilon =1,$ for which all
agents are capable of interacting and converging to the final consensus
value $A.$ Let be $X^{F}=\sum\limits_{\alpha =1}^{N}$ $x_{\alpha }^{F}=NA$
the final total sum of opinions after convergence, where superindexes $I$
and $F$ indicate initial or final states values. From Eq. \ref{uno}, since
in each step $x_{i}(t)+x_{j}(t)=x_{i}(t+1)+x_{j}(t+1)$ is conserved, then
the initial total sum of opinions $X^{I}=\sum\limits_{\alpha =1}^{N}$ $%
x_{\alpha }^{I}$ is equal to $X^{F}.$ Thus the final opinion satisfies $A=%
\frac{1}{N}\sum\limits_{\alpha =1}^{N}$ $x_{\alpha }^{I}$ . When $\epsilon <1
$ we need to resort to a cumbersome double mathematical induction method
over both the number of nodes inside and outside a cluster, but we sketch
here the main ideas. At any step of the simulation, the system may be
broken into subsets because some interactions might be restricted. At any
given time, when \ a node $\ l$ interacts with an element $s$ of a new
cluster with size $M$ (while its former cluster may or not disappear), if
the node $l$ only interacts with the same node $s$ always - a very
improbable case- the proof is trivial, but in general, there is a
competition between diferent clusters that can attract a given node. We
prove in the Appendix how the conservation of opinion holds when this
competition mechanism applies for the simplified case of three nodes aligned
in a simple array. Then, taking into account the fact that in the
competition process opinion is always conserved, we can apply the same
procedure shown above (as if $\epsilon =1$) for all the nodes in the subset
with\ $M+1$ elements.

This conservation law obviously arises from Eq. \ref{uno}, where for all
values of $\mu $ the approach process is symmetric. There is not empirical
evidence that the approach process is exactly symmetric in social systems,
but it could be regarded as a useful rule to simplify simulattions. Two
different random values of $\mu _{i}$ and $\mu _{j}$ in a pair-wise 
interaction is another alternative.

\section{Conclusions}

\bigskip It is important to study directed links in networks because many
real systems relationships and dynamics are not bilateral or reciprocal. 
Here, we first constructed a directed network (DN) from a bidirectional
network UN, by simply deleting half of the links of the UN, to obtain a DN
sharing the same structure. Then we performed simulations within the
Deffuant model of negotiators to find averaged distributions of final
opinions $\left\langle F\right\rangle $ as function of the tolerance $%
\epsilon $ and of the network disorder parameter $p$ in both UN and DN.
Then, by comparing the results of the Deffuant dynamics in these networks we
find a rich structure of the averaged difference of clusters or final
opinions, $\left\langle F_{DN}\right\rangle -\left\langle
F_{UN}\right\rangle .$ The dependence of $\left\langle F\right\rangle $ on $%
\epsilon $ in both UN and DN was much stronger than on $\ p$. Obviously, as
in all bounded confidence models, increasing $\epsilon $ leads to a
monotonous decrease of $\left\langle F\right\rangle $ in each network, but $%
\left\langle F_{DN}\right\rangle -\left\langle F_{UN}\right\rangle,$ 
exhibits a rich structure, since itwas found to be large and positive for
large\ $\epsilon $, but small and negative for smaller values of $\epsilon $%
. The former case follows logically from the fact that, by construction, the
DN have less (half) number of links, as compared to the UN, making the
convergence process to find consensus more difficult in the DN. The presence
of disorder has only a small effect on the convergence time in both UN and
DN, and the larger number of links causes a much faster convergence in
DN. To understand why $\left\langle F_{DN}\right\rangle
-\left\langle F_{UN}\right\rangle $ could be negative, it was helpful to
plot the corresponding distributions of $\ \left\langle F\right\rangle $ \
in the interval [0,1] to find that the DN distributions are less homogeneous
than the analogous  UN distributions. Also for low $\epsilon $ the DN
distributions suffer a relative loss in the number and strength of the
middle opinions, as compared to the UN. Finally, in both UN and DN we show
how final opinions in each cluster are the exact average of the initial
opinions. This conservation property is just a consequence of the two-agent 
symmetrical approach process built into the model. Such  mechanism can
be changed by assigning two different random values for the approach
parameters in each pair-wise interaction without affecting the main results.

We hope that this work may contribute to the advancement of more realistic
models and applications of opinion dynamics in directed networks.

\section{Acknowledgments}

We acknowledge partial financial support from DGAPA-UNAM , M\'{e}xico,
through Grant No IN114208. YG also thanks financial support from Centro
Latinoamericano de F\'{\i}sica (CLAF) in her visit to UNAM, the support from
a graduate fellowship of Misi\'{o}n Ciencia (Venezuela) and the
computational resources kindly provided by the Condensed Matter Physics Lab
at IVIC. We also acknowledge useful discussions and reviewing of this paper by Ernesto
Medina.

\bigskip

\section{Appendix}

If we have three nodes in a row labeled $a$, $b$ and $c$ (with $b$ in the
middle) with initial opinions $a_{0}$, $b_{0}$ and $c_{0}$, we first rewrite
these quantities in terms of the average of the initial opinions:
\begin{eqnarray}
\label{tres}
a_{0} &=\frac{2^{1}+1}{3\left[ 2^{1}\right] }\left\{
a_{0}+b_{0}+c_{0}\right\} -\frac{1}{\left[ 2^{1}\right] }\left\{
-a_{0}+b_{0}+c_{0}\right\}  \\ 
&=\frac{1}{3}\left\{ a_{0}+b_{0}+c_{0}\right\} -\frac{2}{3\left[ 2^{1}%
\right] }\left\{ -2a_{0}+b_{0}+c_{0}\right\},  \nonumber\\
b_{0} &=\frac{2^{1}+1}{3\left[ 2^{1}\right] }\left\{
a_{0}+b_{0}+c_{0}\right\} -\frac{1}{\left[ 2^{1}\right] }\left\{
a_{0}-b_{0}+c_{0}\right\}  \nonumber\\
&=\frac{1}{3}\left\{ a_{0}+b_{0}+c_{0}\right\} -\frac{2}{3\left[ 2^{1}%
\right] }\left\{ a_{0}-2b_{0}+c_{0}\right\},  \nonumber\\
c_{0} &=\frac{2^{1}+1}{3\left[ 2^{1}\right] }\left\{
a_{0}+b_{0}+c_{0}\right\} -\frac{1}{\left[ 2^{1}\right] }\left\{
a_{0}+b_{0}-c_{0}\right\}  \nonumber\\
&=\frac{1}{3}\left\{ a_{0}+b_{0}+c_{0}\right\} -\frac{2}{3\left[ 2^{1}%
\right] }\left\{ a_{0}+b_{0}-2c_{0}\right\}.   
\end{eqnarray}
Then we start the simulation by choosing first node $a$ and its
nearest neighbor node $b$. Here, for simplicity, the approaching parameter $%
\mu $ is $\frac12$, then both opinions $a_{1}$, $b_{1}$ become the same at the middle-point in
this step;
\begin{eqnarray}
\label{cuatro}
a_{1} &=b_{1}=\frac{a_{0}}{2}+\frac{b_{0}}{2} \\
      &=\frac{1}{3}\left\{
a_{0}+b_{0}+c_{0}\right\} -\frac{1}{3\left[ 2^{1}\right] }\left\{
-2a_{0}+b_{0}+c_{0}\right\} -\frac{1}{3\left[ 2^{1}\right] }\left\{
a_{0}-2b_{0}+c_{0}\right\} \\ 
&=\frac{1}{3}\left\{ a_{0}+b_{0}+c_{0}\right\} +\frac{1}{3\left[ 2^{1}%
\right] }\left\{ a_{0}+b_{0}-2c_{0}\right\},   \nonumber\\
c_{1} &=c_{0}=\frac{1}{3}\left\{ a_{0}+b_{0}+c_{0}\right\} -\frac{2}{3\left[
2^{1}\right] }\left\{ a_{0}+b_{0}-2c_{0}\right\} .  
\end{eqnarray}

In the next simulation step, node $c$ and its neighbor $b$ are chosen%
\begin{eqnarray}
\label{cinco}
a_{2} &=a_{1}=\frac{1}{3}\left\{ a_{0}+b_{0}+c_{0}\right\} +\frac{2}{3\left[
2^{2}\right] }\left\{ a_{0}+b_{0}-2c_{0}\right\}  \\ 
b_{2} &=c_{2}=\frac{b_{1}}{2}+\frac{c_{1}}{2}=\frac{1}{3}\left\{
a_{0}+b_{0}+c_{0}\right\} +\left[ 1-2\right] \frac{1}{3\left[ 2^{2}\right] }%
\left\{ a_{0}+b_{0}-2c_{0}\right\}  \nonumber\\
&=\frac{1}{3}\left\{ a_{0}+b_{0}+c_{0}\right\} -\frac{1}{3\left[ 2^{2}%
\right] }\left\{ a_{0}+b_{0}-2c_{0}\right\},  
\end{eqnarray}
to obtain 
\begin{eqnarray}
\label{12}
a_{2n} &=\frac{1}{3}\left\{ a_{0}+b_{0}+c_{0}\right\} +\frac{F}{\left[
2^{2n}\right] }\left\{ a_{0}+b_{0}-2c_{0}\right\}  n=1,2,3,...
\nonumber\\
b_{2n} &=c_{2n}=\frac{1}{3}\left\{ a_{0}+b_{0}+c_{0}\right\} +\frac{G}{%
\left[ 2^{2n}\right] }\left\{ a_{0}+b_{0}-2c_{0}\right\} %
n=1,2,3,... 
\end{eqnarray}%
with $F=\frac{2}{3}$ and \ $G=-\frac{1}{3}$. Notice how rapidly (faster than
a geometrical series) the last terms go to zero  of these equations.

\end{document}